# Universal Numerical Encoder and Profiler Reduces Computing's Memory Wall with Software, FPGA, and SoC Implementations

Al Wegener

*Samplify Systems, Inc.*
*591 W. Hamilton Ave, Suite 250*
*Campbell, CA  95008 USA*
awegener@samplify.com

*Abstract*: In the multi-core era, the time to computational results is increasingly determined by how quickly operands are accessed by cores, rather than by the speed of computation per operand.  From high-performance computing (HPC) to mobile application processors, low multi-core utilization rates result from the slowness of accessing off-chip operands, i.e. the memory wall.  The APplication AXcelerator (APAX) universal numerical encoder reduces computing's memory wall by compressing numerical operands (integers and floats), thereby decreasing CPU access time by 3:1 to 10:1 as operands stream between memory and cores.  APAX encodes numbers using a low-complexity algorithm designed both for time series sensor data and for multi-dimensional data, including images.  APAX encoding parameters are determined by a profiler that quantifies the uncertainty inherent in numerical datasets and recommends encoding parameters reflecting this uncertainty.  Compatible software, FPGA, and system-on-chip (SoC) implementations efficiently support encoding rates between 150 MByte/sec and 1.5 GByte/sec at low power.  On 25 integer and floating-point datasets, we achieved encoding rates between 3:1 and 10:1, with average correlation of 0.999959, while accelerating computational "time to results."

## 1. Motivation & Goals

Numerical computations have accelerated significantly since 2005 thanks to two complementary, silicon-enabled trends:  multi-core processing and single instruction, multiple data (SIMD) accelerators.  As microprocessor designers reached the power limits of ever-increasing CMOS clock rates, they continued to leverage Dennard scaling (Moore's Law) by adding more cores per die, ushering in the multi-core era.  Similarly, Intel's MMX, SSE, and AVX SIMD instruction accelerators generated faster mathematical results by processing multiple operands in parallel with a single instruction.

Unfortunately, due to fundamental limitations of physics, these two trends could not be accompanied by a corresponding increase in memory, storage, and I/O bandwidth.  First coined in 1995 [1], the phrase "memory wall" describes the ever-widening gap between the speed of computational units and the speed of accessing computational operands.  Chip architects try to mitigate the memory wall's effects by using caches that exploit spatial and temporal locality, but doing so only postpones the inevitable.  Additional on-chip cache levels increase overall power consumption and decrease the percentage of silicon area dedicated to computations, such as SIMD. The silicon area of today's leading microprocessors is becoming dominated by increasingly "dark" silicon [2], including some gates in caches that will never be utilized.

High-performance computing (HPC) is the proverbial "canary in the coal mine" of multi-core processing. When HPC hits a fundamental limit, consumer multi-core will likely encounter a similar limit in few years. Twice a year, the HPC community updates its list of the world's fastest supercomputers (www.top500.org). In June 2012 the IBM BlueGene/Q topped the Top500 list by achieving a staggering 16 Petaflops (16 x $10^{15}$ floating-point operations per second [flops]), using 1.57M IBM POWER7 Cores. HPC architects are painfully aware that by 2020, Exaflops ($10^{18}$ flops) may mark the practical end of supercomputers [3], because the annual cost of powering an Exaflops machine will exceed the already substantial cost of buying and assembling its components.

What if the size of numerical operands could be reduced with a more efficient numerical encoding method? The IEEE-754 (2008) floating-point standard [4] describes both computational and storage formats, so the idea of a dedicated numerical storage format is not new. The APplication AXcelerator (APAX) numerical encoder is flexible, adaptable, efficient, and implemented in a small number of CMOS gates. APAX can reduce computing's memory wall in a scalable, low-power way. Understanding why the numerical encoding method works requires a deeper investigation into the characteristics of computational datasets and an acknowledgement of computational uncertainty. We propose that numerical operands be encoded during multi-core block writes (processor transfers to DDR/disk/network) and decoded during multi-core block reads (DDR/disk/network transfers to processors). The proposed encoding format increases both the capacity and the bandwidth of all computer components surrounding multi-core sockets: DDR memory; PCIe, Ethernet, and Infiniband networks; and solid state memory and disk drives. APAX achieves encoding rates between 3:1 and 10:1 without changing the statistical or spectral characteristics of decoded datasets. At the profiler-selected encoding rate, our method preserves the results of every computational application our collaborators have examined. Figure 1 illustrates how memory-bound compute applications achieve 1.8x to 3x better CPU efficiency and faster "time to results."

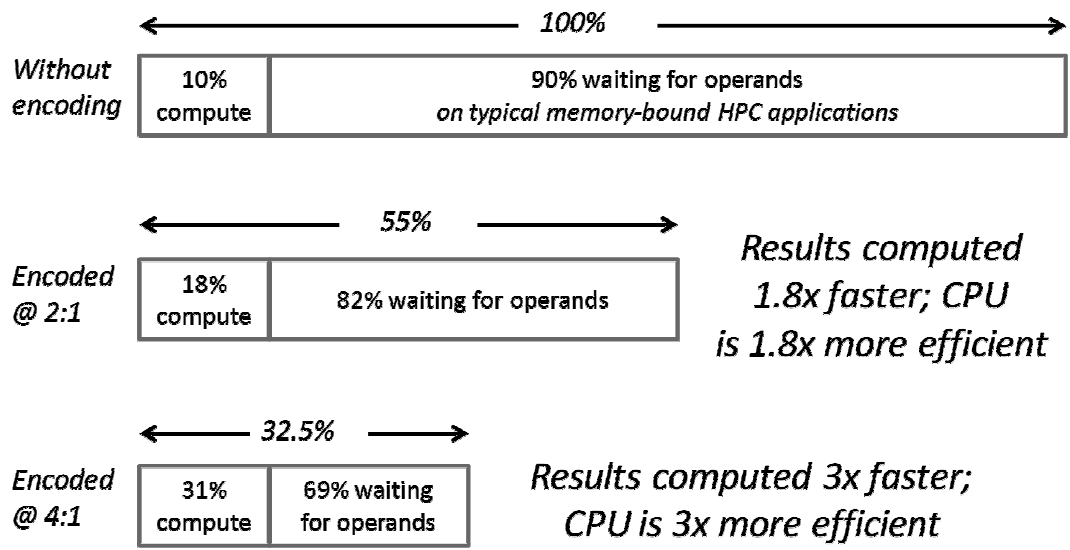

**Figure 1:** Numerical encoding at 2:1 (4:1) shortens "time to results" by 1.8x (3x)

## 2. Discussion of Prior Art

Numerical compression has an exceedingly rich history dating back to the 1960s. However, the compression research community has understandably focused on compression of consumer speech, audio, images, and video, given the ubiquity of such media. But media compression techniques are inappropriate for a universal numerical encoder because the quality metrics for media compression are determined by limitations of human hearing and vision, not by the accuracy requirements of numerical computations. Columnar compression [5] for numbers is only useful when numbers are stored in databases. Compressive sensing (CS) [6, 7] is targeted towards real-time integer sensor systems but requires significant back-end complexity to unwind the analog front end's random sampling. Waveform coders [8, 9] encode integer samples but can lack flexibility. Lossless compression of scientific data [10-13] comes closest to the numerical encoding method we describe here but typically achieves less than a 2:1 encoding ratio on floats. Portending the low-power future, inexact CMOS computing circuits have recently been described [14], and even Intel has developed an approximate, "good enough" low-power multiplier [15]. Our work distinguishes over the prior art by allowing users to flexibly encode both integers and floats at an operating point on each dataset's unique rate-distortion curve, as recommended by the accompany profiler. For 32-bit single-precision floats, the profiler's recommended operating point maintains "five nines" (0.99999) of correlation with the original dataset. By combining the numerical encoder with a profiler, scientists can discover and explore dataset uncertainty in a way that both informs and surprises. APAX encoding combines flexibility of numeric representation with algorithmic efficiency in both software & hardware.

## 3. Computational Uncertainty

HPC computations simulate a noisy analog world. Sensors provide the input to many HPC simulations. If not used directly as HPC input, sensor data is used to validate HPC results and predictions. Yet in many HPC simulations, integer sensor data is over-cast into 32-bit or 64-bit floating-point numbers for computational convenience. For example, when 16-bit integers are cast to single-precision floats and all downstream computations are performed using single precision operations, over-casting masks uncertainties in the original data and engenders a false sense of confidence in the accuracy of results. Sensors employ imperfect analog-to-digital converters (ADCs) whose effective number of bits (ENOB) per sample is always less than the ADC's resolution, often by 2 or more bits. ADC resolution is a datasheet bullet, but ENOB is the definitive ADC accuracy metric. HPC simulations regularly use 32-bit floats that were converted from (or compare HPC results to) an ADC's 16-bit integers, which may only have 14-bit ENOB. The APAX profiler exposes this over-casting of HPC arrays by estimating dataset uncertainty using a variety of metrics, including signal spectra.

Sensitivity analysis and uncertainty quantification (UQ) estimate the uncertainty in HPC input and intermediate data and carry it thru HPC calculations to final results. UQ, an extension of the "significant digits" methodology that most of use learned in high school science, quantifies confidence bounds for HPC results. Unfortunately UQ software makes HPC codes run at least 3x slower, making UQ's widespread adoption unlikely in HPC. In contrast, frequency-domain UQ spectral estimation techniques [16] provide a consistent, proven, and low-MIPS way to quantify the uncertainty of both integers and floats. Since most sensor signals are both bandlimited and cyclostationary, uncertainty can be spectrally characterized for entire classes of signals and need not operate continuously.

Let us now consider uncertainties inherent in HPC calculations. Broadly studied but infrequently acknowledged, floating-point math is inexact and approximate [17]. Paradoxically, compiler optimizations that make HPC codes run faster often change floating point results by re-ordering instructions. Some compiler directives that claim to simply replace single-core floating-point unit (FPU) instructions with their SIMD-"equivalent" instructions also change results. The recently approved EEMBC FPmark floating-point benchmark gives passing grades to floating-point codes whose results, compared to "gold standard" floating-point results, only match 14 of 23, or 31 of 52, mantissa bits. Such practical decisions by industry experts acknowledge that compiler optimizations may create *multiple* "correct" results, all of which may be *good enough* for each benchmark code's intended computational purpose.

Finally, let's reflect upon the nature of HPC results and Big Data analytics. If "the purpose of computing is insight, not numbers," (R. W. Hamming), it seems reasonable that humans will more quickly gain insight from a GB result than a TB result. In Big Data parlance, this is the difference between business data and business intelligence. HPC simulations reduce TB of floating-point input data to GB or MB of "results" that, ironically, are often visualized on an 8-bit RGB monitor. HPC researchers sometimes puzzlingly insist that every one of their TB input floats must be losslessly encoded while being unable to explain how negligible, zero-mean changes to individual samples could possibly affect the much smaller computational result. When HPC results are orders of magnitude smaller than the datasets from which they are created, some sort of filtering (averaging) must be at the heart of such computations. Most filtering is linear and thus preserves signal statistics and spectral content while reducing input data to a more manageable, but still massive, scale. In communications systems, this intentional and desirable data reduction reduces incoherent noise and is called processing gain.

In summary, numerical computations are subject to multiple sources of uncertainty that scientists often ignore for expediency and to avoid complexity. Would it not be better to estimate this uncertainty while flexibly encoding numbers to reduce computing's memory wall? As long as the encoding method preserves dataset statistics and spectral content, our users' experience indicates that computational results will be *good enough* for the intended use, because "good enough" is controlled by the encoder user. HPC and mobile computing regularly make resolution vs. speed tradeoffs. Since the degree of uncertainty introduced by APAX is demonstrably smaller than these other sources of uncertainty, APAX should be a welcome addition to engineers' tradeoff toolbox.

## 4. APAX Profiler

The APAX profiler is a software utility that creates a rate-correlation graph, provides 18 quantitative metrics comparing the original and decoded signals, recommends an encoding rate, and displays two signal quality graphs. Figure 2 illustrates the APAX-profiled results of an HPC climate variable rsds.nc (NetCDF file). Profiler results appear in four windows: 1. rate-correlation window, 2. results summary window, 3. spectral window, and 4. signal-residual distribution window.

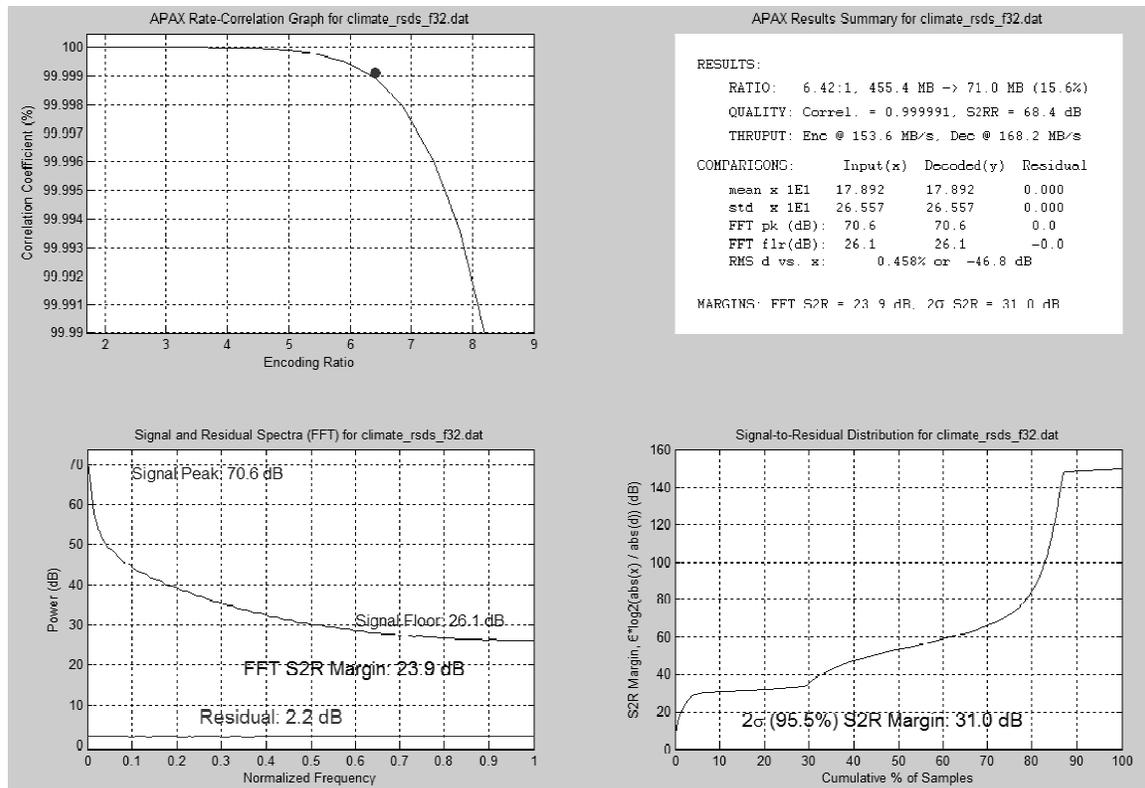

**Figure 2:** APAX profiler results on a 32-bit climate dataset (rsds.nc)

Profiler Window 1 displays the relationship between the APAX encoding rate and the resulting Pearson's correlation coefficient. The profiler recommends the operating point (the dot on the rate-correlation graph) where the correlation between the original data x and the decoded data y is 0.99999 ("five nines"). After the profiler's results are displayed at the recommended operating point, profiler users can drag & drop the operating point to a different encoding rate or a different correlation setting on the rate-correlation curve. With each new operating point, the profiler updates Windows 2 to 4. The Pearson's correlation coefficient $r$ quantifies the similarity between input (x) and decoded (y) samples:

$$r = \frac{\sum xy}{\sqrt{\sum x^2 \sum y^2}}$$

Profiler Window 2 displays 18 quantitative similarity and difference metrics comparing sampled input x(i), decoded output y(i), and residual or difference d(i) = x(i)-y(i) signals. Signal quality is expressed both as correlation coefficient r and as the signal-to-residual ratio (SRR), further described below. Larger r and SRR values indicate better decoded signal quality. Using a single x86 core, APAX encoding and decoding speeds are between 130 MB/sec (encode) and 170 MB/sec (decode). Window 2 also compares the means, standard deviations, spectral peaks, and noise floors of x, y, and d. Two RMS metrics compare the magnitude of residual d to the magnitude of signal x, in both percent and dB. Finally, Window 2 summarizes frequency-domain margin FFT S2R and time-domain margin 2σ S2R (see below). The larger the margin, the smaller the residual.

Profiler Window 3 compares the input signal's spectrum (upper curve), to the spectrum of the residual signal (lower curve). For the HPC climate variable profiled in Figure 2, the input signal's spectral peak is 71 dB and its floor is 26 dB, indicating a dynamic range (DR) of 45 dB. DR is inversely proportional to the uncertainty of the input signal; the larger the DR, the lower the uncertainty. Importantly, the residual spectrum has a mean value of 2 dB, which is *nearly 24 dB below the noise floor of the input signal!* APAX encoding-decoding generates a spectrally flat (white) residual spectrum, so the encoding's effects are comparable to adding low-amplitude white noise to input signal x, where APAX users control the noise amplitude. The FFT S2R margin is the difference between input signal floor (26 dB) and residual mean spectral value (2 dB). The SRR is the difference between the input spectral peak and the residual mean. Users control FFT S2R margin and SRR by selecting the operating point in Window 1.

Profiler Window 4 graphs the signal-to-residual (S2R) distribution and calculates the $2\sigma$ (95.5%) S2R margin. The S2R distribution is calculated by integrating the SRR probability distribution. For the climate variable in Figure 2 at an encoding rate of 6.42:1, the $2\sigma$ S2R margin is 31 dB, meaning that 95.5% of APAX-decoded samples $y(i)$ have residuals $d(i)$ whose magnitude is at least 31 dB smaller than corresponding input sample $x(i)$. Users control $2\sigma$ S2R margin by selecting the operating point in Window 1.

## 5. Signal Characteristics Exploited by APAX Encoding

As illustrated in Figure 3, APAX encoding takes advantage of three characteristics of all numerical sequences: peak-to-average ratio (Fig 3a), oversampling (Fig. 3b), and effective bits (Fig. 3c). Figure 3a illustrates that a signal's peak value can be much larger than its average value. The peak-to-average ratio (PAR) of many integer and floating-point datasets often exceeds 10 dB. APAX's entropy encoder uses a block floating-point format to indicate how many bits are encoded in each group of 4 samples, thus tracking signal magnitude. Figure 3b's example 3G wireless signal spectrum (10 MHz bandwidth sampled at 60 Msamp/sec) illustrates that numerical sequences are often oversampled, causing sample-to-sample redundancy in the time domain. The higher the oversampling ratio, the greater the redundancy. APAX encoding reduces sample-to-sample correlations by encoding selected derivatives of input datasets. Figure 3c illustrates the distinction between an ADC's resolution and its ENOB. ENOB becomes obscured when integers are cast to floating-point numbers; the profiler reveals ENOB and quantifies uncertainty. For both integer and floating-point values, ENOB affects dataset uncertainty and thus numerical accuracy. By controlling decoded dataset accuracy, APAX encoding preserves the bits that matter while reducing each dataset's bit rate, thus lowering the memory wall.

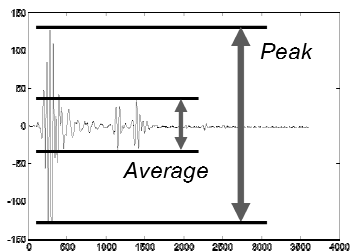
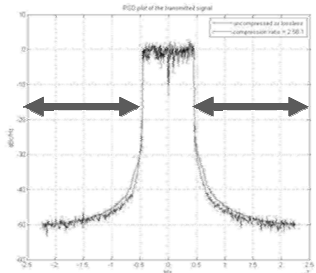
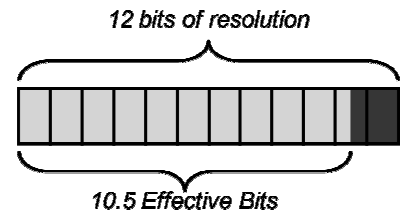

**Figure 3a:** Peak-to-Average Ratio

**Figure 3b:** Oversampling

**Figure 3c:** Effective Number of Bits

## 6. APAX Encoder: Block Diagram and Implementations

Figure 4 presents a block diagram of the APAX Encoder, whose six components are described below. Additional details are available in [18]. The APAX algorithm encodes sequential blocks of input data elements with user-selected block size between 64 and 16,384 (any multiple of 4). The signal monitor tracks various characteristics of the input dataset, including center frequency and SNR. The center frequency estimate determines which nearby data elements of (possibly modulated) waveforms are correlated and also estimates dataset uncertainty. The attenuator multiplies each sample of the input dataset by a floating-point value that can vary from block to block, under the control of an adaptive loop that converges to the user's desired encoding rate (compressed packet size) or correlation target. The block header generator assembles parameters from the signal monitor, the redundancy remover, and the encoding mode into a 4-Byte (ints) or 6-Byte (floats) header that precedes each block's encoded numerical payload. The redundancy remover generates filtered versions of the attenuated data series and determines which of the numeric streams (the attenuated data elements themselves or their filtered versions) encodes using the fewest bits. The "best stream" decision for block j is encoded in the header of block j+1. The APAX control block orchestrates and synchronizes overall encoder operation.

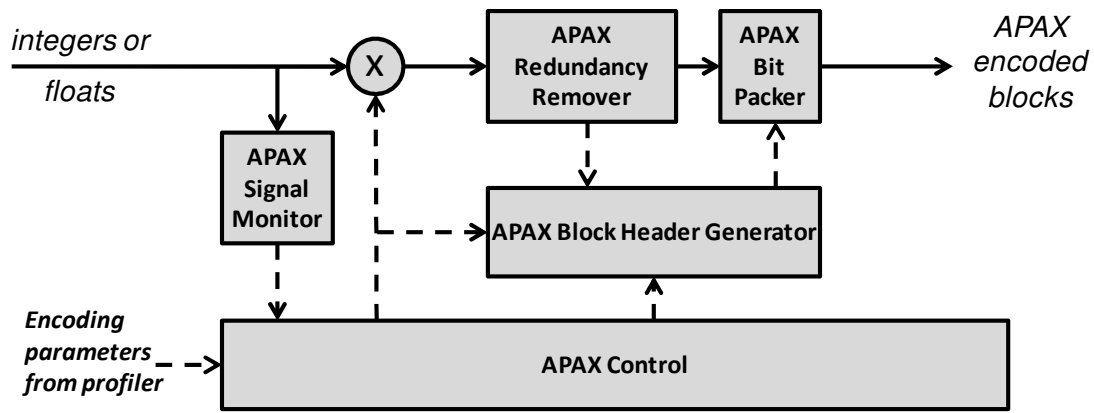

**Figure 4:** APAX Encoder Block Diagram

The bit packer (entropy encoder), further described in [19], encodes groups of 4 or 8 successive samples by modifying a well-known bit-packing technique called block floating point encoding (BFPE). Earlier BFPE methods encode an exponent (exp) that indicates the number of bits in each of the next N samples. For APAX, N=4. Our modification to BFPE, called joint exponent encoding (JEE), exploits the fact that BFPE exponents are highly correlated. Successive BFPE exp differences (diffs) fall in the range {-1,+1} about 80% of the time. When two successive exp diffs are both in the range {-1,+1}, JEE sends one of nine 4-bit tokens. When successive exp diffs cannot be paired and each diff falls in the {-2, +2} range, JEE encodes each exp diff using one of five 4-bit tokens. All remaining exponents are absolutely encoded using an 8-bit token. JEE encoding efficiency can exceed Huffman and Golomb-Rice entropy efficiency because JEE tokens typically represent two exp diffs. JEE tokens are fixed-length, hardware-friendly 4-bit or 8-bit values. The first exponent of each APAX block is absolutely encoded. APAX encoders and decoders are implemented in compatible software and hardware products. Table 1 summarizes the software, FPGA, and silicon complexity (resource usage) and corresponding thruput of each APAX implementation.

| Implementation | Complexity | Thruput |
|---|---|---|
| Software | ~50 asm instructions per sample | 150 MB/sec per x86 core |
| FPGA netlist | Enc: ~4,000 6-input LUTs<br>Dec: ~3,500 6-input LUTs | 600 MB/sec per instance |
| SoC IP block | Enc + dec: ~220k gates<br>( < 0.1 $mm^2$ in 28 nm CMOS) | 1.5 GB/sec per instance |

**Table 1:** APAX Resource Utilization

## 7. Results

The 25 numerical datasets summarized in Table 2, including 8-bit, 16-bit, and 32-bit ints and 32-bit and 64-bit floats, demonstrate APAX's encoding performance. The datasets represent a variety of HPC application areas, including climate, geology, multi-physics, communications, scientific visualization, medical imaging, test & measurement, and embedded systems. The application variety illustrates that the signal characteristics described in Section 5 are present in all of these numerical sequences. The largest data files contain climate variables from NetCDF files (32-bit floats). The 32-bit integers datasets contain bandlimited signals with varying center frequency, bandwidth (spectral occupancy), and SNR. The multi-physics signals represent characteristics of plasma heated by lasers. The 64-bit double precision data files were taken from various scientific applications [11]. As indicated in Table 2 by each dataset's spectral peak, 32-bit integers have the largest magnitudes while multi-physics floats contain the smallest magnitudes. Interestingly, the dynamic range of these datasets ranges from 23 dB (64-bit floats) to 89 dB (32-bit floats). While 32-bit floats can represent numbers across 76 orders of magnitude (from $10^{-38}$ to $10^{+38}$), scientific datasets rarely require more than 6 orders of magnitude (about 20 bits), which is consistent with our premise that many scientific datasets are over-cast.

Table 3 illustrates the encoding ratios and signal quality results when Table 2's datasets are encoded at the profiler's recommended operating point and then decoded. Across all 25 datasets, the average encoding ratio is 6.43:1 and the average Pearson's correlation coefficient is 0.999959 ("four nines").

| Input Dataset | Data Type | Size (MB) | Spectral peak (dB) | Spectral floor (dB) | Dyn Rng (dB) | Input Dataset | Data Type | Size (MB) | Spectral peak (dB) | Spectral floor (dB) | Dyn Rng (dB) |
|---|---|---|---|---|---|---|---|---|---|---|---|
| image1_i8.dat | int8 | 37.75 | 58.7 | -4.0 | 62.7 | climate_ta500_f32.dat | single | 227.72 | 70.1 | -18.7 | 88.8 |
| image2_i8.dat | int8 | 9.44 | 62.8 | 13.0 | 49.8 | geology1_f32.dat | single | 80.00 | 8.3 | -37.3 | 45.6 |
| oscilloscope_i8.dat | int8 | 2.00 | 48.3 | -8.2 | 56.5 | geology2_f32.dat | single | 60.83 | -35.5 | -64.5 | 29.1 |
| CT_50views_i16.dat | int16 | 5.84 | 107.6 | 71.2 | 36.4 | multiphys_den_f32.dat | single | 4.79 | -113.2 | -141.0 | 27.7 |
| image3_i16.dat | int16 | 75.50 | 98.5 | 37.8 | 60.7 | multiphys_rho_f32.dat | single | 4.92 | 5.2 | -67.6 | 72.8 |
| bandlim1_i32.dat | int32 | 4.00 | 133.6 | 75.5 | 58.1 | multiphys_t0_f32.dat | single | 4.79 | -30.8 | -62.5 | 31.7 |
| bandlim2_i32.dat | int32 | 4.00 | 107.2 | 56.4 | 50.8 | multiphys_t2_f32.dat | single | 4.79 | -22.9 | -58.6 | 35.7 |
| bandlim3_i32.dat | int32 | 4.00 | 106.9 | 60.1 | 46.8 | | | | | | |
| bandlim4_i32.dat | int32 | 4.00 | 115.4 | 66.8 | 48.6 | msg_bt_f64.dat | double | 12.00 | 73.9 | 17.6 | 56.3 |
| bandlim5_i32.dat | int32 | 4.00 | 110.7 | 68.0 | 42.7 | num_brain_f64.dat | double | 12.00 | 19.1 | -31.8 | 50.9 |
| climate_huss_f32.dat | single | 455.45 | -17.4 | -74.1 | 56.7 | num_control_f64.dat | double | 12.00 | -31.1 | -53.9 | 22.8 |
| climate_pr_f32.dat | single | 455.45 | -64.8 | -92.7 | 27.9 | num_plasma_f64.dat | double | 12.00 | -29.3 | -77.0 | 47.7 |
| climate_rsds_f32.dat | single | 455.45 | 70.6 | 26.1 | 44.5 | obs_temp_f64.dat | double | 12.00 | 23.7 | -2.2 | 26.0 |

**Table 2:** Input Datasets

| Input Dataset | Encoding Ratio | InSize (MB) | EncSize (MB) | Correlation | SDR (dB) | RMS resid (%) | RMS resid (dB) | Spectral Margin (dB) | 2σ S2R Margin (dB) |
|---|---|---|---|---|---|---|---|---|---|
| image1_i8.dat | 2.83 | 37.75 | 13.34 | 0.999167 | 61.3 | 1.28% | -37.9 | 0.5 | 25.1 |
| image3_i8.dat | 3.92 | 9.44 | 2.41 | 1.000000 | 52.0 | 5.01% | -26 | 2.2 | 12.2 |
| oscilloscope_i8.dat | 2.02 | 2.00 | 0.99 | 0.999264 | 40.4 | 4.90% | -26.1 | 3.7 | 15.3 |
| CT_50views_i16.dat | 7.92 | 5.84 | 0.74 | 0.999272 | 51.9 | 3.91% | -28.15 | 15.6 | 21.0 |
| image2_i16.dat | 4.52 | 75.50 | 16.70 | 0.999981 | 65.1 | 0.68% | -43.4 | 4.4 | 22.8 |
| bandlim1_i32.dat | 4.56 | 4.00 | 0.88 | 1.000000 | 62.7 | 0.62% | -44.11 | 4.6 | 27.4 |
| bandlim2_i32.dat | 3.91 | 4.00 | 1.02 | 0.999994 | 55.5 | 0.34% | -49.4 | 4.7 | 27.4 |
| bandlim3_i32.dat | 3.74 | 4.00 | 1.07 | 0.999995 | 51.2 | 0.47% | -46.63 | 4.3 | 24.3 |
| bandlim4_i32.dat | 2.95 | 4.00 | 1.35 | 1.000000 | 77.6 | 0.04% | -67.01 | 29.0 | 45.6 |
| bandlim5_i32.dat | 3.90 | 4.00 | 1.03 | 1.000000 | 56.2 | 0.37% | -48.63 | 13.5 | 27.4 |
| climate_huss_f32.dat | 9.21 | 455.45 | 49.43 | 0.999975 | 65.0 | 0.77% | -42.29 | 8.3 | 33.3 |
| climate_pr_f32.dat | 10.66 | 455.45 | 42.72 | 0.999992 | 62.0 | 0.42% | -47.44 | 34.1 | 29.1 |
| climate_rsds_f32.dat | 6.42 | 455.45 | 70.99 | 0.999991 | 68.4 | 0.46% | -46.79 | 23.9 | 31.0 |
| climate_ta500_f32.dat | 9.81 | 227.72 | 23.22 | 1.000000 | 95.9 | 0.02% | -73.01 | 7.1 | 68.1 |
| geology1_f32.dat | 5.06 | 80.00 | 15.82 | 0.999978 | 47.3 | 0.73% | -42.68 | 1.7 | 20.5 |
| geology2_f32.dat | 7.48 | 60.83 | 8.14 | 0.999995 | 78.2 | 0.46% | -46.66 | 49.1 | 41.8 |
| multiphys_den_f32.dat | 2.88 | 4.79 | 1.66 | 1.000000 | 64.2 | 0.09% | -60.91 | 36.5 | 39.5 |
| multiphys_rho_f32.dat | 8.50 | 4.92 | 0.58 | 1.000000 | 110.7 | 0.00% | -87.83 | 37.9 | 82.9 |
| multiphys_t0_f32.dat | 2.91 | 4.79 | 1.64 | 0.999999 | 69.5 | 0.08% | -61.6 | 37.8 | 39.5 |
| multiphys_t2_f32.dat | 3.78 | 4.79 | 1.27 | 0.999997 | 56.7 | 0.45% | -46.94 | 21.0 | 25.8 |
| msg_bt_f64.dat | 10.19 | 12.00 | 1.18 | 0.999996 | 84.4 | 0.08% | -61.84 | 28.1 | 18.2 |
| num_brain_f64.dat | 10.63 | 12.00 | 1.13 | 1.000000 | 91.6 | 0.04% | -69 | 40.7 | 63.5 |
| num_control_f64.dat | 7.77 | 12.00 | 1.54 | 0.999993 | 52.8 | 0.43% | -47.25 | 30.0 | 21.3 |
| num_plasma_f64.dat | 7.77 | 12.00 | 1.54 | 0.999958 | 65.2 | 0.63% | -44.07 | 17.5 | 19.9 |
| obs_temp_f64.dat | 6.94 | 12.00 | 1.73 | 0.999990 | 65.9 | 0.34% | -49.28 | 40.0 | 24.3 |
| **AVERAGES:** | **6.43** | | | **0.999959** | **68.1** | **0.52%** | **-52.50** | **22.25** | **34.30** |

**Table 3:** APAX Encoding Results

Significantly, the average spectral margin (difference between input signal noise floor and average residual power) is *more than 22 dB*, and the average 2σ S2R margin exceeds 34 dB. Validation performed by HPC collaborators who provided the climate, multi-physics, and geology datasets (anonymous due to confidentiality agreements), confirmed that their HPC simulation results were unchanged when APAX-decoded datasets were used in place of the original datasets. The results shown in Table 3 were obtained at software thruput rates between 130 MB/sec (encode) and 170 MB/sec (decode).

APAX thruput (MB/sec) increases with datatype width. Encoding and decoding 64-bit floats is faster than encoding and decoding 8-bit integers because APAX encodes samples, not Bytes. In contrast, APAX hardware thruput is independent of datatype and exceeds APAX software thruput by factors between 4x (FPGA) and 10x (SoC IP block).

We welcome and encourage collaboration with other data scientists on other numerical datasets that can further substantiate the degree to which APAX encoding enables faster "time to results" without changing those results.

## 8. Conclusion

We have described a universal numerical encoder and profiler that reduces computing's memory wall in a user-controlled way. The profiler recommends an operating point that users can modify for each profiled numerical dataset. For memory-bound applications, encoder software can use some of the processor's idle compute cycles to decode previously encoded datasets arriving from disk drives, SSDs, DDR memory, or even L3 cache. As an SoC IP block, the encoder enhances DDR, DMA, and I/O controllers by encoding numerical data as it leaves multi-core sockets and decoding such data when it arrives. The APAX IP block trades pins for gates, accelerating multi-core's limited I/O resources using a flexible, adaptive, and efficient algorithm. Adding APAX as an SoC IP block to silicon die that already include multiple cores and SIMD accelerators is well justified, because APAX encoding accelerates time to results.